# An adaptive hybrid algorithm for social networks to choose groups with independent members

Parham Hadikhani[*1], Pooria Hadikhani[2]

**Abstract**

Choosing a committee with independent members in social networks can be named as a problem in group selection and independence in the committee is considered as the main criterion of this selection. Independence is calculated based on the social distance between group members. Although there are many solutions to solve the problem of group selection in social networks, such as selection of the target set or community detection, just one solution has been proposed to choose committee members based on their independence as a measure of group performance. In this paper, a new adaptive hybrid algorithm is proposed to select the best committee members to maximize the independence of the committees. This algorithm is a combination of particle swarm optimization (PSO) algorithm with two local search algorithms. The goal of this work is to combine exploration and exploitation to improve the efficiency of the proposed algorithm and obtain the optimal solution. Additionally, to combine local search algorithms with particle swarm optimization, an effective selection mechanism is used to select a suitable local search algorithm to combine with particle swarm optimization during the search process. The results of experimental simulation are compared with the well-known and successful metaheuristic algorithms. This comparison shows that the proposed method improves the group independence by at least 21%.

## 1 Introduction

Nowadays, due to the presence of the internet, social networks are considered as an integral part of human life. Social networks are a new generation of websites that have become the focus of attention of internet users throughout the world these days. Social networks have a social structure formed by sets of individuals (or organizations or other social entities). These are related to social relations such as information exchange, cooperation, friendship, relationship, or financial transactions [1]. Moreover, the social network has become a subject of research in many different disciplines, parallel to the continued growth of the internet, which allows cooperation and collaboration between people. Research in a number of academic fields has shown that social networks are applied in many levels, which play an important role in determining issues, managing organizations and how successful people are in achieving their goals. Therefore, many organizations use social networks to find people for a committee who can make decisions about specific problems. Recognition of groups is a popular subject in computer science. The structure of a group is made up of the people interactions. Selected people among the group are expected to decide in a way that benefits the entire group and avoid the closeness and intimacy that exists between them. In this context, the best committees are those who show the greatest independence among their members, which can be achieved by defining criteria among members of a group. An issue that has been examined is Committee Selection Problem. In this issue the objective is to choose the people in social who have maximum independence among others to form a committee [1]. The independence is based on distances between each pair of committee member. The greater the distance between committee members, the greater the

Parham Hadikhani
parhamhadikhani@gmail.com

[1]Department of Computer Engineering, Pasargad Higher Education Institute, Shiraz, Fars, Iran
[2]School of Mechanical Engineering, College of Engineering University of Tehran, Tehran, Iran

independence between committee members. The figure 1 is a salient example for this problem. In this paper, we propose a hybrid algorithm to solve this problem. In this algorithm, we hybridize the binary particle swarm optimization algorithm with two local search algorithm including simulated annealing and hill climbing. The particle swarm optimization algorithm is a population-based approach and many population-based approaches are not good at exploiting the areas around the explored solutions [2]. Therefore, to solve this problem, a local search algorithm is hybridized to improve the efficiency of particle swarm optimization algorithm as the local search algorithms has a good exploitation performance [2]. Additionally, to combine local search algorithms with particle swarm optimization, an effective selection mechanism [2] is used to select a suitable local search algorithm to combine with particle swarm optimization during the search process.

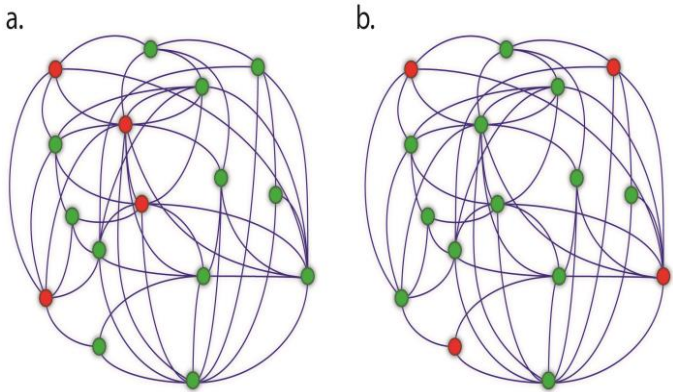

Figure 1: this is a sample of the show for selecting the committee. (a) Indicates the committee members are close and their independence is low. However, (b) reveals committee members that have great independence among them because of their distance

Choosing committees with independent members on social networks is one of the group selection problem [1]. Group selection problem can be divided in two well-known categories [1]. First category is the target set selection problem. The goal of this problem is finding nodes that have great influence to increase and distribute something like information. The second one is the community detection problem. In this problem, the goal is to discover groups of nodes which have similar relations and properties. Regarding the target set selection problem, in [3], they developed the genetic algorithm to improve the optimal solution. In [4], they presented a theoretical method to find influence overlap. Afterwards, to remove this overlap, they proposed two algorithms of DRS and DSN. Furthermore, considering the literature of community detection, in [5], they proposed a novel modularity function to improve the community detection in social networks. In [6], they provided a new approach for community detection in social networks using leader nodes which their algorithm has two steps. First, they detect the leaders. After that, they detect the community based on their similarity. Also in [7], a new algorithm was proposed to remove mostly random selection and reduces the iteration times and keeps the original time efficiency.

## 2 Discrete Particle Swarm Optimization Algorithm

The particle swarm optimization algorithm was proposed in 1995 by James Kennedy and Russell Eberhat [8].it made a strong algorithm was created in the field of optimization to improve the optimal solutions. The basis of this algorithm is inspired by the social behavior of the groups of birds seeking food. A group of birds in the search space are randomly looking for food. There is only one piece of food in the search space. Each solution called a particle in this algorithm which is equivalent to a bird in the movement of birds. Each particle has a fitness value that is calculated by a fitness function. The closer the particle in the search space to the target (food in the bird's motion model), the more appropriate it is. Each particle also has a speed that guides the movement of the particle. Each particle continues to move through the search space by following the optimum particles in the current state. How this algorithm works is that a group of particles are created at random at the start of the work and try to find an optimal solution by updating generations. At each step, each particle is updated using the two best values. The first is the best situation the particle has ever achieved called $P_{best}$. The next best value used by the algorithm is the best situation ever achieved by the population of particles called $P_{g\,.best}$. After finding the best values, the velocity and the location of each particle are updated using the equations (1) and (2):

$$1)\ V_i(t+1) = w * V_i(t) + c_1 * rand_1 * (P_{i\,.best} - X_i(t)) + c_2 * rand_2 * (P_{g\,.best} - X_i(t))$$

$$2)\ X_i(t+1) = X_i(t) + V_i(t+1)$$

The above method is proposed in the continuous optimization environment and its application in the discrete space of our problem is not feasible and for this reason, we use discrete particle swarm optimization method.

The discrete PSO algorithm is introduced with BinaryPSO (BPSO). The search space in BPSO is considered as a hypercube in which a particle may be seen to move to nearer and farther corners of the hypercube by flipping various numbers of bits [9]. The moving velocity is defined in terms of changes of probabilities that a bit will be in one state or the other. Thus a particle moves in a state-space restricted to 0 and 1 on each dimension, where each id v represents the probability of bit id x taking the value 1. With this definition, id p and id x are integers in {0, 1} and $v_{id}$, since it is a probability, it must be constrained to the interval of [0.0, 1.0] [10]. By defining a logistic function transformation $S(v_{id})$ [11] in equation (3), the position will be updated according to equation (4).

1) $S(v_{id}) = sigmoid(v_{id}) = \frac{1}{1+e^{-v_{id}}}$

2) $if\ rand(\ ) < S(v_{id})\ then\ x_{id}(t+1) = 1$
   $else\quad x_{id}(t+1) = 0$

In the above equation, $S(v_{id})$ is a sigmoid limiting transformation and rand () is a quasi-random number selected from a uniform distribution of [0.1, 1.0]. In the continuous version of PSO, id v is limited by the value max v. Also in the discrete version of PSO, id v is limited in the range of [- max v, max v] [12]. Usually, max v is set to 6. Although this setting limits the probability to be in the range of [0.0025 0.9975] but it will be resulted in a better convergence characteristic. It should be noted that as standard PSO, the BPSO can be implemented through global and local models. In this paper, both models are used.

## 3 Proposed method

It is possible to improve the results of different optimization problems by combining the local search algorithms with the population based algorithms. The local search algorithms are good at exploitation and the population based algorithms are powerful in extraction [11]. Therefore, the combination of these algorithms can benefit from their strength simultaneously and provides an efficient search algorithm. Considering the successful implementation of hybrid algorithms in the literature and the current improvements of the population based search algorithms, we propose a new hybrid algorithm which is a combination of BPSO and a local search algorithm. There are different local search algorithms in the literature where they have different approaches for escaping from the local minima. These algorithms are suitable for special cases or specific search steps. Thus, it is challenging to determine the best local search algorithm that has the best performance in combination with BPSO in various conditions. We propose an adaptive BPSO algorithm that resolves the problem of choosing the best local search algorithm to be combined with the BPSO. The adaptive BPSO algorithm uses a compatible selection mechanism for choosing an appropriate local search algorithm that can be combined with the BPSO. This selection is based on the fitness function improvement of local search algorithms that is obtained after running the local search algorithms on the current particle. The proposed adaptive BPSO operates as follows (Figure 2):

- Considering a set of local search algorithms and the amount of their fitness, the selection mechanism is called to select one of the local search algorithms to apply on the generated answer.
- We update the amount of fitness improvement of the selected algorithm and check its completion criterion.
- If the algorithm has achieved the completion criterion, the search is finished and the best solution is stored. Otherwise, the above process is repeated.

In the following sections, the local serach algorithms and selection mechanism is discussed.

## 4 Local search algorithms

Local search algorithms are used in combination with population-based algorithms like PSO to speed up the convergence process. Most of the local search algorithms starts with an initial solution and then repeatedly explores the surrounding area [13]. If the solution is better than the first solution, it keeps the solution and then it continues by searching in a new area to promote the solution. The discovery of an adjacent area is achieved by a neighbor. The neighborhood discovery process is repeated until it

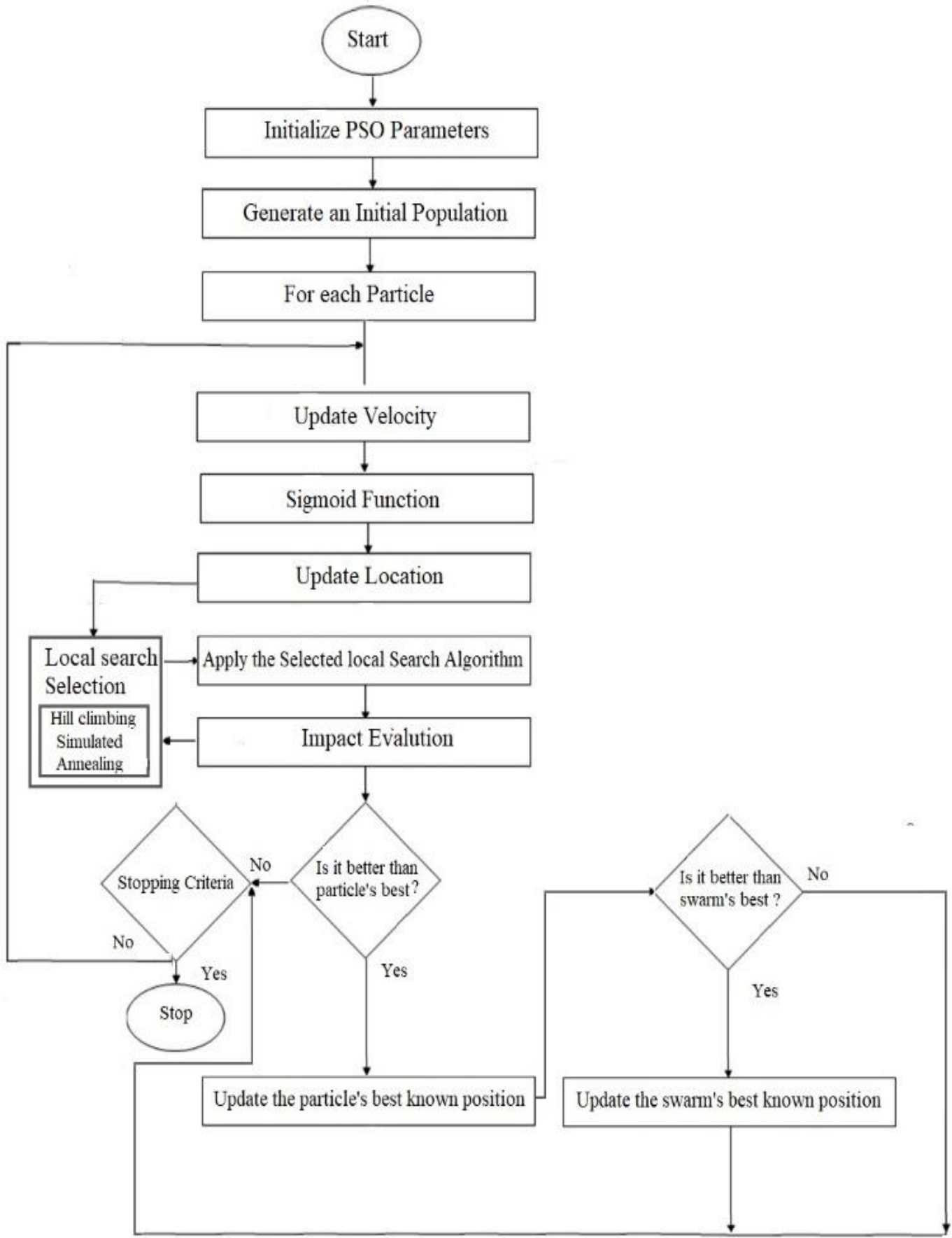

Figure 2: The proposed algorithm

## 4.1 Hill climbing algorithm

The Hill climbing algorithm is an optimization technique belonging to the family of local search algorithms. It is an iterative technique that starts with an arbitrary solution and then tries to achieve a better solution by changing an element of the solution. If this change leads to a better solution, another change will be made to this new solution. This process will continue until there is no further improvement in the solution [14].

## 4.2 Simulated Annealing

This algorithm was invented in 1983 by Scott Kirkpatrick and Daniel Gelatt [15] and is based on the process of metal annealing. In the annealing process, the metals are first heated to very high temperatures, and then, the process of cooling and lowering the temperature gradually takes place on them. In this process, as the heat of the metal increases, the movement speed of its atoms is greatly increased, and in the next step, the gradual decrease in temperature results in the formation of specific patterns in the placement of its atoms. This change in the pattern of atoms gives rise to valuable properties in the refrigerant metal, including its increased strength. To solve an optimization problem, the SA algorithm first starts with an initial solution and then moves to a neighboring solution in an iteration loop. If the neighbor solution is better than the current answer, the algorithm sets it as the current answer, otherwise the algorithm accepts the answer with the probability of exp(-ΔE / T) as the current answer. In this equation, ΔE is the difference between the objective function of the current solution and the neighbor solution and T is a parameter called temperature. At each temperature, several repetitions are performed and then the temperature is slowly lowered. Initial steps set the temperature too high to be more likely to accept worse responses. With the gradual decrease in the temperature, the final steps are less likely to accept worse solutions, and as a result, the algorithm converges to a good solution.

## 5 The adaptive selection mechanism

The method described in [2,16] is used to choose the best local search algorithm for the problem. Initially, at each iteration, a reward (credit) $r_{ls}$ is assigned to the selected local search and it is calculated using the solution quality of the local search at the current iteration as follows [1]:

$$1)\ r_{ls} = \frac{\left(f(S)-f(S')\right)}{|f(S)|}$$

Where $f(S)$ is the cost of the current particle which is obtained from BPSO and $f(S')$ is the cost of the particle derived from selected local search algorithm.

Afterwards, to evaluate the local search strategies, an empirical quality estimate $q_{ls}$ is defined which determines the average reward achieved by the selected local search until the current iteration. The quality estimate of the local search is updated at each iteration by the following equation [2].

$$2)\ q_{ls.i} = q_{ls.i-1} + \frac{1}{n_{ls.i-1}}(r_{ls} - q_{ls.i-1})$$

In the above equation, $n_{ls}$ is the number of times that $ls$ algorithm is selected, and $i$ and $i-1$ subscripts denote the current and the previous iteration, respectively.

Finally, Multi-Armed Bandit algorithm (MAB) [17] is utilized to automatically choose the most suitable local search. MAB chooses the best algorithm considering the quality of an algorithm and the number of times that the algorithm is selected. MAS is formulated as follows:

$$3)\ Selected\ LS = \arg\max_{ls=1...K}\left(q_{ls.i} + C\sqrt{\frac{\log\sum_1^K n_{k.i}}{n_{ls.i}}}\right)$$

Where $K$ is the number of local search strategies. Variable $C$ in the above equation is a scaling factor that creates a balance between selecting a local search algorithm with the best empirical quality estimate and selecting a rarely chosen local search algorithm. Algorithm 1 shows the pseudo code of the adaptive selection of local search algorithms. In this algorithm, the initial value of all variables is zero. The details are described in the Algorithm1.

- input : Current particle
- output: Selected algorithm and new particle

If the algorithms have not been applied yet then

    Each local search algorithm must be applied at least once

Else use the following formulated to select one of them:

$$Selected\ LS = \arg\max_{ls=1\ldots K}\left(q_{ls.i} + C\sqrt{\frac{\log\sum_1^K n_{k.i}}{n_{ls.i}}}\right)$$

End if

After selecting, the $q_{ls}$, $r_{ls}$, and the number of times selected local search is updated.

End

Algorithm 1: Adaptive Selection Mechanism

## 6 Fitness function

To calculate the fitness of the proposed algorithm, we use the fitness function that is proposed in the [1]. This function works based on the distance and the purpose of this function is finding the longest distance between each pair of nodes to reach an independent committee. In order to obtain an independent committee, this function uses the size of the committee and the network diameter (c and D respectively). In the following, the equation of the Fitness function is given:

$$1)\ f = \frac{\left[\sum_{i,j=0}^{k} s(i,j)/c\right] + L}{2*D}$$

In this equation s is the distance between each pair of nodes and L is the lowest distance among the members of committee.

## 7 Result and discussion

In this section, we present the simulation results of the proposed algorithm and compare its results with well-known algorithms which include, the simulated annealing, hill climbing, genetic, and particle swarm optimization. For each algorithm, between five and ten simulation steps were performed and the average of their outputs are reported in the graphs. Moreover, we use Facebook data set which is downloaded from Stanford Large Network Dataset for Collection. In the Following, first, we analyze Facebook social network. Next, the parameters set-up and configuration of our algorithm are discussed. Finally, the results of the proposed algorithm with others are analyzed.

### 7.1 Data Set

The graph on Facebook's social network has a very high density. The general information about the Facebook social network is found in the Table1. As expected, this graph has a large clustering coefficient and the average shortest path. Additionally, in the figure 3 the graph shows the degree distribution of the Facebook's graph. People who active on Facebook will have more edges than those not. A few people have a large number of degrees and the majority of people have small number of degrees.

Table 1: Facebook Social Network Information

| Average shortest path | Diameter | Clustering coefficient | Average degree graph | Number of edges | Number of vertices | Graph |
|---|---|---|---|---|---|---|
| 3/693 | 8 | 0/605 | 43/6910 | 88234 | 4039 | facebook.com |

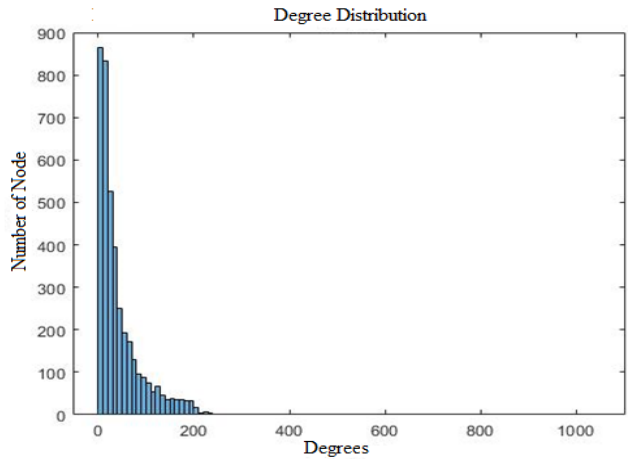

Figure 3: Degree Distribution of Facebook

## 7.2 Parameters set-up

A preliminary test [18] is performed to find the appropriate values for the parameters of each algorithm that is used in the proposed algorithm. Initially, we randomly selected a set of parameters. All algorithms presented in the proposed algorithm were tested with different parameter values. The best values were selected after the 10 runs. The following sections, we discuss the parameters of the algorithms used in the proposed algorithm:

### 7.2.1 PSO parameter settings

The particle swarm algorithm has five parameters. The first parameter called the inertia coefficient (w) indicates the particle tendency to maintain the current state of motion. $c_1$ and $c_2$ are the learning coefficients. $c_1$ is The personal learning coefficient and $c_2$ is the collective learning coefficient. In our work, These parameters have been fixed to w=2, $c_1$=2 and $c_2$=2 . Moreover, the lower and upper limits of the variables are equal to zero and one respectively.

### 7.2.2 Selection mechanism parameters settings

The selection mechanism has a parameter which is the scaling factor (C). In this paper we set it to 0.01 which is showed in Table 2.

### 7.2.3 Hill Climbing parameter settings

In the hill climbing algorithm used there is only one parameter that is the Maximum Iteration that corresponds to the number of iterations in the hill climbing algorithm. This parameter is set to 3000 based on several examination which have been done on Maximum Iteration values.

### 7.2.4 Simulated Annealing parameter settings

The simulated annealing that we used in this paper, has three parameters. These parameters are including: the maximum temperature ($T_{max}$), the minimum temperature ($T_{min}$) and the cooling rate (β). In our work, $T_{min}$, $T_{max}$ and β were fixed to 1, 1000 and 0.99, respectively.

Table 2: Summary of the parameters used in the proposed algorithm

| Parameters | Algorithm | Value |
|---|---|---|
| Maximum Iteration | Hill Climbing | 3000 |
| $T_{max}$ | Simulated Annealing | 1 |
| $T_{min}$ | Simulated Annealing | 1000 |
| β | Simulated Annealing | 0.99 |
| $c_1$ | PSO | 2 |
| $c_2$ | PSO | 2 |
| w | PSO | 2 |
| $V_{min}$ | PSO | 0 |
| $V_{max}$ | PSO | 1 |
| C | MAB | 0.01 |

We developed a new algorithm for evaluating the groups of actors with the greatest distance between them, which are assumed to be independent criteria. This configuration has the following parameters:

• Size of the community: If we consider p as the number of people in the social network and n as the number of people needed for the committee, the size of the committees is obtained by p / n.
• Stop criterion: If the iteration of the proposed algorithm is more than the specified value, the algorithm stops.
• Runs: 40 runs are generated by 5 runs per configuration.

## 7.3 Parameters evaluation

Parameters are examined to evaluate the effectiveness of the proposed method:
- Centrality criteria
- 3-member, 4-member, and 5-member committees

## 7.4 Social network metrics

The social network criteria for the current committees are shown in Table 3. The average degree in the network is 43/6910 and network diameter is 8. The mean path length is also 5.4320. From Table 3, by increasing the number of committee members, the degree of nodes which correspond with people, increases to a value higher than average value. In addition, some committee members (such as A3) show very few betweenness, but the closeness between each of them is more balanced.

Table 2:3-member, 4-member, and 5-member committees with each degree, betweenness, and closeness

| Committees | Nodes | Degree | Betweenness | Closeness |
|---|---|---|---|---|
| 3-member | A1 | 22 | 7/8982 | 4/4207 |
|  | A2 | 18 | 270/7009 | 6/4408 |
| fitness=0/8125 | A3 | 1 | 0 | 4/5562 |
| 4-member | R1 | 7 | 3/22619 | 6/4687 |
|  | R2 | 43 | 31/6318 | 4/4230 |
| fitness=0/8281 | R3 | 17 | 10/53416 | 7/0028 |
|  | R4 | 14 | 12/92619 | 4/5589 |
| 5-member | F1 | 69 | 287/1839 | 4/4281 |
|  | F2 | 84 | 206/3849 | 7/8623 |
|  | F3 | 68 | 3555/1117 | 8/3022 |
| fitness=0/9625 | F4 | 126 | 576/0439 | 7/1240 |
|  | F5 | 28 | 13/1543 | 7/0082 |

## 7.5 simulation of member committees

Figures 4 to 6 illustrate the results of experimental simulation of 3-member, 4-member, and 5-member committees for the proposed algorithm in comparison with the genetic algorithm, particle swarm optimization, simulated annealing, and hill climbing. As it can be seen in Figs. 4-1, the proposed algorithm has a significant advantage over the aforementioned algorithms. The reason for this superiority is the combination of local search algorithms (simulated annealing and hill climbing) with binary particle swarm optimization algorithm using an intelligent mechanism to select the appropriate local search algorithm in each run of the algorithm. Additionally, since the proposed algorithm uses the combination of Exploration and Exploitation simultaneously with respect to Fitness Function, it can find the best possible optimum solution compared to the other mentioned algorithms. Simulated annealing and hill climbing algorithms also have the worst results among the other algorithms. The reason for this is that they do not have the capability of Global search to extract. Moreover, although genetic and particle swarm optimization algorithms have been able to find a relatively optimal solution, they have not been able to find the best optimal solution as it is clear from the figures.

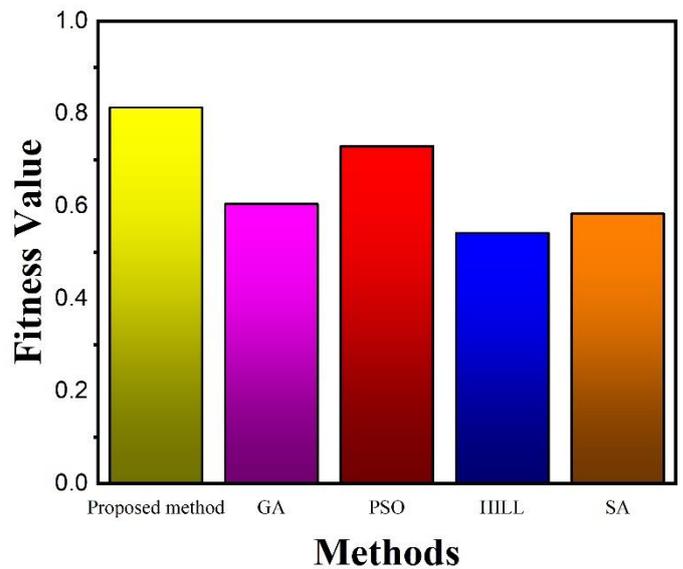

Figure 4:3-member committee

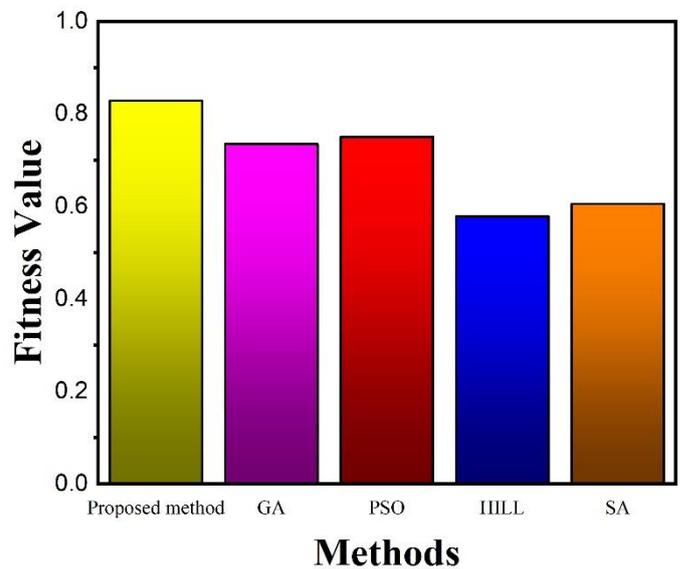

Figure 5:4-member committee

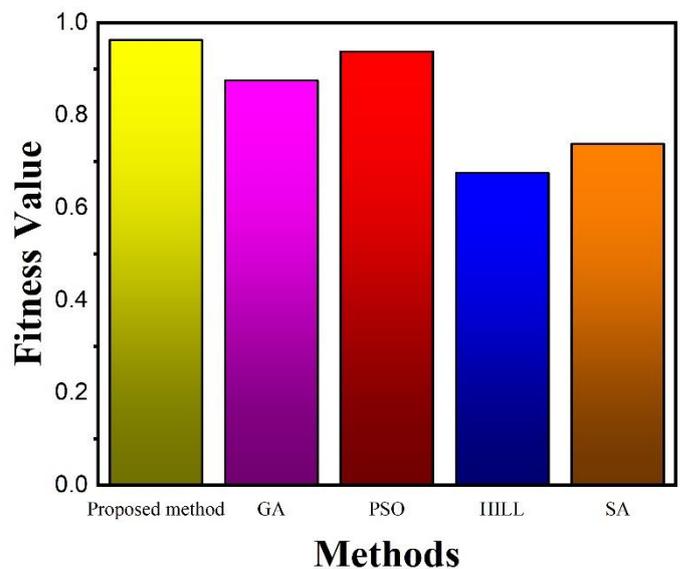

Figure 6:5-member committee

Figure 7 also shows the comparison of the 3-member committee diagram among the proposed algorithm, Genetic algorithm, and PSO.

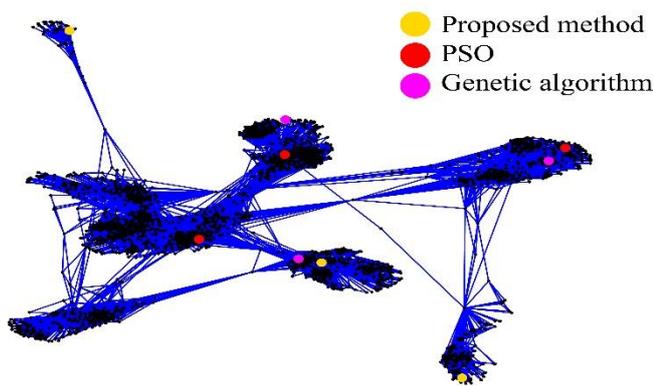

Figure 7:3-member committee diagram among the proposed algorithm, Genetic algorithm, and PSO

## 8 Conclusion

In this paper, we propose an adaptive hybrid algorithm which combines the PSO algorithm and two local search algorithms of SA and hill climbing with an adaptive selection mechanism to select the appropriate local search algorithm in each iteration. The role of selection mechanism is to select a local search algorithm that gives the best result to combine it with the PSO. The proposed algorithm has been presented to the committee member selection problem in social networks whose members are independent. In this issue of group selection for the selection of committee members, independence is the main criterion of choice for the independent function of the group. This group independence function uses geodesic distances to measure the social distance between each pair of nodes in social networks. Also, our proposed algorithm is applied to guarantee the best committee candidates. The independent group function is then maximized to select the candidate groups with the best fitness. The results show that the performance of the particle swarm optimization algorithm is enhanced by combining it with a local search algorithm. Because PSO algorithm has a shortcoming of converging prematurely after getting trapped into some local optima (local optimum solution point) and considers it to be the global optima (global optimum solution point), its combination with the local search algorithm creates a balance between exploration and exploitation and guarantee the global convergence.